\documentstyle[preprint,aps]{revtex}

\tighten
\begin{document}

\title{
Large-basis shell-model calculations for p-shell nuclei
}
\medskip

\author{
        P. Navr\'atil\footnote{On leave of absence from the
   Institute of Nuclear Physics,
                   Academy of Sciences of the Czech Republic,
                   250 68 \v{R}e\v{z} near Prague,
                     Czech Republic.}
        and B. R. Barrett
        }

\medskip

\address{
Department of Physics, University of Arizona, Tucson, 
Arizona 85721 
}

\maketitle

\bigskip

\begin{abstract}
Results of large-basis shell-model calculations for nuclei
with $A=7-11$ are presented. The effective interactions used in the study 
were derived microscopically
from the Reid93 potential and take into account the Coulomb potential
as well as the charge dependence of $T=1$ partial waves.  
For $A=7$, a $6\hbar\Omega$ model space was used, while for the rest 
of the studied nuclides, the calculations were performed in a 
$4\hbar\Omega$ model space. It is demonstrated that the shell model 
combined with microscopic effective interactions derived from modern 
nucleon-nucleon potentials is capable of providing good agreement 
with the experimental properties of the ground state as well as 
with those of the low-lying excited states.
\end{abstract}

\bigskip
\bigskip
\bigskip

\narrowtext



\section{Introduction}
\label{sec1}

Large-basis no-core shell-model calculations have recently been 
performed 
\cite{ZBJVC,JZBV,JHBV,ZBVC,ZB94,ZVB,ZBVM,ZBVHS,r:NB96,r:NTB97,r:NBOFermi}. 
In these calculations all nucleons are active,
which simplifies the effective interaction as no hole states
are present. In the approach taken, the effective interaction 
derived microscopically from modern nucleon-nucleon potentials is
determined for a system of two nucleons only and subsequently
used in the many-particle calculations. To take into account a
part of the many-body effects a, so-called, multi-valued effective 
interaction approach was introduced and applied in the no-core 
shell-model calculations \cite{ZBVHS}. 

In the past, these calculations concentrated on the $0s$-shell
nuclei and $A=5,6$ $0p$-shell nuclei. In addition, $^{7}$Li was studied
in a $4\hbar\Omega$ model-space calculation \cite{ZBVHS}
and the obtained wave-functions were employed for evaluating proton
and electron scattering characteristics \cite{KBAD97}.  
Large-basis no-core shell-model calculations for heavier 
$0p$-shell nuclei have not been discussed until recently.
In the most recent application, we applied the no-core shell-model
approach to the $A=10$ nuclei in order to evaluate the isospin-mixing
correction to the Fermi matrix element 
$^{10}$C$\rightarrow ^{10}$B \cite{r:NBOFermi}. In the present paper
we complement this study by presenting the results for other light 
$0p$-shell nuclides. In particular, we present calculations for the 
$A=7$ nuclei
$^7$He, $^7$Li, $^7$Be and $^7$B performed in a $6\hbar\Omega$ model space,
in which configurations up to an energy of $6\hbar\Omega$ relative 
to the unperturbed ground-state configuration are included.
For $A=8$, results were obtained for $^8$He, $^8$Li, $^8$Be and $^8$B
in a $4\hbar\Omega$ model space. For $A=9$ we calculated properties
of $^9$He, $^9$Li, $^9$Be, $^9$B and $^9$C also in a $4\hbar\Omega$ 
model space.
As the $A=10$ nuclides $^{10}$C, $^{10}$B, and partly $^{10}$Be, were 
discussed in Ref. \cite{r:NBOFermi}, 
we now complete the $A=10$ nuclei description by including results 
for $^{10}$He,
$^{10}$Li and $^{10}$Be. Finally, we give the results of a $4\hbar\Omega$ 
calculation for $^{11}$Li and $^{11}$Be.

Our study is distinguished
from other $0p$-shell-nuclei calculations by the fact that we are using
microscopically derived effective interactions, contrary to phenomenological
interactions employed in most other papers \cite{CK65,K74,HWG88,WB92,PWG93}. 
The other distinguishing factor is the use of large multi-configuration 
model spaces.

The organization for the paper is as follows. First, in 
Section~\ref{sec2} we discuss the shell-model Hamiltonian
with a bound center-of-mass and 
the method used to derive the starting-energy-independent
effective interaction. 
Results of the calculations for $A=7-11$  
are presented in Section \ref{sec3} and 
concluding remarks are given in Section \ref{sec4}.

\section{The shell-model Hamiltonian and effective interaction}
\label{sec2}

In the present paper we apply the approach discussed
in Refs. \cite{r:NB96,r:NBOFermi}.
We start with the one- plus two-body Hamiltonian for the A-nucleon system, 
i.e.,
\begin{equation}\label{ham}
H^\Omega=\sum_{i=1}^A \frac{\vec{p}_i^2}{2m}+\sum_{i<j}^A 
V_{\rm N}(\vec{r}_i-\vec{r}_j) + \frac{1}{2}Am\Omega^2 \vec{R}^2\ ,
\end{equation}
where $m$ is the nucleon mass, $V_{\rm N}(\vec{r}_i-\vec{r}_j)$ 
the nucleon-nucleon interaction
and $\frac{1}{2}Am\Omega^2 \vec{R}^2$ 
$(\vec{R}=\frac{1}{A}\sum_{i=1}^{A}\vec{r}_i)$ is the
the center-of-mass harmonic-oscillator potential.
The latter potential does not influence intrinsic properties of the 
many-body system. It provides, however, a mean field felt by each nucleon
and allows us to work with a convenient harmonic-oscillator basis.
The Hamiltonian (\ref{ham}), depending on the harmonic-oscillator 
frequency $\Omega$, may be cast into the form
\begin{equation}\label{hamomega}
H^\Omega=\sum_{i=1}^A \left[ \frac{\vec{p}_i^2}{2m}
+\frac{1}{2}m\Omega^2 \vec{r}^2_i
\right] + \sum_{i<j}^A \left[ V_{\rm N}(\vec{r}_i-\vec{r}_j)
-\frac{m\Omega^2}{2A}
(\vec{r}_i-\vec{r}_j)^2
\right] \; .
\end{equation}
The one-body term of the Hamiltonian (\ref{hamomega}) is then re-written
as a sum of the center-of-mass term
$H^\Omega_{\rm cm}=\frac{\vec{P}_{\rm cm}^2}{2Am}
+\frac{1}{2}Am\Omega^2 \vec{R}^2$,
$\vec{P}_{\rm cm}=\sum_{i=1}^A \vec{p}_i$,
and a term depending only on relative coordinates.
Shell-model calculations are carried out in a model space defined
by a projector $P$. In the present work, we will always use a complete 
$N\hbar\Omega$ model space which includes all the configurations 
up to an energy of $N\hbar\Omega$ relative 
to the unperturbed ground-state configuration.
The complementary space to the model space
is defined by the projector $Q=1-P$.
In addition, from among the eigenstates of the Hamiltonian 
(\ref{hamomega}),
it is necessary to choose only those corresponding to the same 
center-of-mass energy. This can be achieved by projecting 
the center-of-mass eigenstates
with energies greater than $\frac{3}{2}\hbar\Omega$ upwards in the
energy spectrum. The shell-model Hamiltonian, used in the actual 
calculations, takes the form
\begin{eqnarray}\label{phamomegabeta}
H^\Omega_{P\beta}=\sum_{i<j=1}^A &P&\left[ 
\frac{(\vec{p}_i-\vec{p}_j)^2}{2Am}
+\frac{m\Omega^2}{2A} (\vec{r}_i-\vec{r}_j)^2
\right]P + \sum_{i<j}^A P\left[ V_{ij}-\frac{m\Omega^2}{2A}
(\vec{r}_i-\vec{r}_j)^2
\right]_{\rm eff} P  \nonumber \\
&+& \beta P(H^\Omega_{\rm cm}-\frac{3}{2}\hbar\Omega)P\; ,
\end{eqnarray}
where $\beta$ is a sufficiently large positive parameter.
In Eq. (\ref{phamomegabeta}), the notation $[\;\;]_{\rm eff}$ 
means that the quantity within the square brackets is the residual
interaction to be used in the determination of the effective
interaction within the model space $P$ (see Ref. \cite{r:NB96} for more
details).

The effective interaction introduced in Eq. (\ref{phamomegabeta})
should, in principle, exactly reproduce the full-space results
in the model space for some subset of states.
In practice, the effective interactions can
never be calculated exactly, because, in general, for an $A$-nucleon system 
an $A$-body effective interaction is required. 
Consequently, large model spaces
are desirable when only an approximate effective interaction is used. 
In that case, the calculation
should be less affected by any imprecision of the effective
interaction. 
The same is true for the evaluation of any observable characterized
by an operator. In the model space, renormalized effective operators 
are also required. The larger the model space, the less renormalization
is needed.

Usually, the effective interaction is approximated by a 
two-body effective interaction determined from a two-nucleon
system. In this study, we use the procedure, as described in 
Ref.~\cite{r:NB96}.   
To construct the effective interaction we employ
the Lee-Suzuki \cite{r:LS80} similarity transformation
method, which gives an interaction in the form
$P_2 V_{\rm eff}P_2 = P_2 V P_2 + P_2V Q_2\omega P_2$,
with $\omega$ the transformation operator satisfying $\omega=Q_2 \omega P_2$.
The projection operators $P_2, Q_2=1-P_2$ project on the two-nucleon
model and complementary spaces, respectively.
Note that we distinguish the two-nucleon system projection operators
$P_2, Q_2$ from the A-nucleon system operators $P, Q$.
Our calculations start with exact solutions of the Hamiltonian
\begin{equation}\label{hamomega2}
H^\Omega_2\equiv H^\Omega_{02}+V_2^\Omega=
\frac{\vec{p}_1^2+\vec{p}_2^2}{2m}
+\frac{1}{2}m\Omega^2 (\vec{r}^2_1+\vec{r}^2_2)
+ V(\vec{r}_1-\vec{r}_2)-\frac{m\Omega^2}{2A}(\vec{r}_1-\vec{r}_2)^2 \; ,
\end{equation}
which is the shell-model Hamiltonian (\ref{hamomega}) applied to
a two-nucleon system. 
We construct the effective interaction directly
from these solutions. Let us denote the two-nucleon 
harmonic-oscillator states, which form the model space, 
as $|\alpha_P\rangle$,
and those which belong to the Q-space, as $|\alpha_Q\rangle$.
Then the Q-space components of an eigenvector $|k\rangle$ of
the Hamiltonian (\ref{hamomega2}) can be expressed as a combination
of the P-space components with the help of the operator $\omega$
\begin{equation}\label{eigomega}  
\langle\alpha_Q|k\rangle=\sum_{\alpha_P}
\langle\alpha_Q|\omega|\alpha_P\rangle \langle\alpha_P|k\rangle \; .
\end{equation}
If the dimension of the model space is $d_P$, we may choose a set
${\cal K}$ of $d_P$ eigenevectors, typically the lowest states
obtained in each channel, 
for which the relation (\ref{eigomega}) 
will be satisfied. Under the condition that the $d_P\times d_P$ 
matrix $\langle\alpha_P|k\rangle$ for $|k\rangle\in{\cal K}$
is invertible, the operator $\omega$ can be determined from 
(\ref{eigomega}).   
Once the operator $\omega$ is determined the effective Hamiltonian
can be constructed as follows 
\begin{equation}\label{effomega}
\langle \gamma_P|H_{2\rm eff}|\alpha_P\rangle =\sum_{k\in{\cal K}}
\left[
\langle\gamma_P|k\rangle E_k\langle k|\alpha_P\rangle
+\sum_{\alpha_Q}\langle\gamma_P|k\rangle E_k\langle k|\alpha_Q\rangle
\langle\alpha_Q |\omega|\alpha_P\rangle\right] \; .
\end{equation}
This Hamiltonian, when diagonalized in a model-space basis, reproduces
exactly the set ${\cal K}$ of $d_P$ eigenvalues $E_k$. Note that
the effective Hamiltonian is, in general, quasi-Hermitian. 
It can be hermitized by a similarity transformation 
determined from the metric operator $P_2(1+\omega^\dagger\omega)P_2$. 
The Hermitian Hamiltonian is then given by \cite{r:S82SO83}
\begin{equation}\label{hermeffomega}
\bar{H}_{\rm 2eff}
=\left[P_2(1+\omega^\dagger\omega)P_2\right]^{1/2}
H_{\rm 2eff}\left[P_2(1+\omega^\dagger\omega)
P_2\right]^{-1/2} \; .
\end{equation}

Finally, the two-body effective interaction used 
in the present calculations
is determined from the two-nucleon effective Hamiltonian 
(\ref{hermeffomega}) as $V_{\rm eff}=\bar{H}_{\rm 2eff}-H_{02}^\Omega$.

To at least partially take into account the many-body effects
neglected when using only a two-body effective interaction,
we employ the recently introduced, so-called, multi-valued effective
interaction approach \cite{ZBVHS}. In that approach, 
different effective interactions are used 
for different harmonic-oscillator excitations of the spectators.
The effective interactions then carry an additional index 
indicating the sum of the oscillator quanta for the spectators,
$N_{\rm sps}$, defined by
\begin{equation}\label{Nsps}
N_{\rm sps} = N_{\rm sum} - N_{\alpha} - N_{\rm spsmin}
= N'_{\rm sum} - N_{\gamma} - N_{\rm spsmin} \; ,
\end{equation}
where $N_{\rm sum}$ and $N'_{\rm sum}$ are the total oscillator
quanta in the initial and final many-body states, respectively, 
and $N_{\alpha}$
and $N_{\gamma}$ are the total oscillator quanta in the initial
and final two-nucleon states $|\alpha\rangle$ and $|\gamma\rangle$,
respectively. $N_{\rm spsmin}$ is the minimal value of the spectator
harmonic-oscillator quanta for a given system. E.g., for A=7, 
$N_{\rm spsmin}=1$.
Different sets of the effective interaction are determined
for different model spaces characterized by $N_{\rm sps}$ 
and defined by projection operators
\begin{mathletters}\label{projop}\begin{eqnarray}
Q_2(N_{\rm sps})&=&\left\{ 
\begin{array}{ll}
0 &  \mbox{if  $N_1+N_2\leq N_{\rm max} - N_{\rm sps}$} \; , \\ 
1 &  \mbox{otherwise} \; ;
\end{array}
\right.   \\
P_2(N_{\rm sps}) &=&
1-Q_2(N_{\rm sps}) \; .
\end{eqnarray}\end{mathletters}
In Eqs. (\ref{projop}), $N_{\rm max}$ characterizes the
two-nucleon model space. It is an input parameter chosen in relation
to the size of the many-nucleon model space.
This multi-valued effective-interaction approach is superior
to the traditional effective interaction,
as confirmed also in a model calculation \cite{r:NB96l}.

\section{Application to the p-shell nuclei}
\label{sec3}

We apply the formalism outlined in section \ref{sec2} for 
selected $0p$-shell nuclei. 
In the calculations we use the Reid93 nucleon-nucleon potential \cite{r:SKTS}
and consider the following isospin-breaking contributions. First, the Reid93
potential differs in the $T=1$ channels for proton-neutron (pn) and 
proton-proton (pp), neutron-neutron (nn) systems, respectively.
Second, we add the Coulomb potential to the pp Reid93 potential. Consequently,
using the Eqs. (\ref{eigomega})-(\ref{hermeffomega}),
we derive different two-body effective interactions for the pn, pp, and
nn systems. 

As we derive the effective interaction microscopically from the nucleon-nucleon
interaction, the number of freely adjustable  parameters in the calculation 
is limited.
First, we have the choice of the model-space size in the shell-model 
diagonalization.
That is, however, constrained by computer capabilities. The largest model space
we were able to use was the space allowing all $6\hbar\Omega$ excitations 
relative to the unperturbed ground state
for $A=7$ nuclei and all $4\hbar\Omega$ excitations 
relative to the unperturbed ground state
for $A>7$ nuclei, respectively. The calculations were done in the 
m-scheme using the Many-Fermion-Dynamics
Code \cite{r:VZ94} extended to allow the use of different pn, pp, nn 
interactions. We note that in this study the same effective 
interaction is used for each isobaric chain, and thus the same model-space
size is employed for all the isobars of given $A$.

Second, our effective interactions depend on the choice of the two-nucleon 
model space size. The two-nucleon model space size is related to the 
many-nucleon model-space size, and, in principle, is determined by that size.
Traditionally, however, the $Q_2=0$ space used to determine the G-matrix
does not, neccesarily, coincide with the many-particle model space 
\cite{r:BHM71,r:HJKO95}.
In our calculation, the two-nucleon model space is characterized by 
a restriction on the number of harmonic-oscillator quanta   
$N_1\le N_{\rm max}$, $N_2\le N_{\rm max}$, $(N_1+N_2)\le N_{\rm max}$.
Here, $N_i=2n_i+l_i$ is the harmonic-oscillator quantum number for the nucleon
$i, i=1,2$. This type of restriction guarantees an orthogonal transformation
between the two-particle states and the relative- and center-of-mass-coordinate
states. With regard to the $4\hbar\Omega$ calculation for the $0p$-shell nuclei, 
the choice of $N_{\rm max}=6$ appears to be appropriate.
However, it has been observed in the past~\cite{ZBVC,ZBVM,r:NB96} that 
when the Lee-Suzuki
procedure is combined with the G-matrix calculation according to 
Ref.~\cite{r:BHM71} (which is equivalent to the procedure we are using)
and is applied to calculate the two-body effective interaction,
the resulting interaction may be too strong. This is, in particular, true,
when the multi-valued approach is used.  
The reason for this is likely the fact that our effective interaction
is computed for a two-nucleon system bound in an harmonic-oscillator
potential. Therefore, artificial binding from this potential is included
in the effective interaction and the many-body effects coming from 
the large-basis space calculation do not completely compensate 
for this spurious binding.
This effect decreases when the model-space 
size increases as is demonstrated in our earlier three-nucleon shell-model
calculations \cite{r:NBthree}.  
Several possible adjustments have been discussed to deal with this problem 
\cite{ZBVC,r:NB96} in smaller model spaces
and amount to introducing an extra parameter. In the present calculations,
we use two methods introduced in the previous papers.
For the $A=10$ $4\hbar\Omega$ calculation presented in Ref.
\cite{r:NBOFermi}, we preferred 
to treat $N_{\rm max}$ as a free parameter and use $N_{\rm max}=8$
for the $4\hbar\Omega$ calculations. With this choice, which results
in an overall weaker interaction than that calculated with 
$N_{\rm max}=6$, we obtain
quite reasonable binding energies for all the studied nuclei. This
approach is employed here for the $4\hbar\Omega$ calculations
for the nuclei with $A>7$.  
On the other hand, in the case of $A=7$ for the $6\hbar\Omega$ space 
we follow Ref. \cite{r:NB96}
and use the parameter $k_Q$ introduced there. While for
$k_Q=1$ there is no modification, the choice of $k_Q<1$ reduces the 
contribution of the $Q_2$-space part of the harmonic-oscillator potential
on the two-nucleon effective interaction. For this calculation, 
$N_{\rm max}=8$
appropriate for the $6\hbar\Omega$ many-nucleon space size is employed.
In Ref. \cite{r:NB96} we used this approach for the 
$6\hbar\Omega$ calculations for $A=5$ and 6 nuclei as well as for
the $8\hbar\Omega$ calculations for $A=3$ and 4 nuclei.  
We note that we are studying ways on how to eliminate these extra
parameters by means of renormalizations of the two-body effective 
interactions utilizing knowledge of the three-body effective interactions.
The three-body effective interactions can be calculated following
the approach introduced in Ref. \cite{r:NBthree}.

Finally, our results depend on the harmonic-oscillator frequency $\Omega$.
We have studied this dependence by performing calculations
for the values $\hbar\Omega=14, 15.5$, and $17$ MeV for $^{10}$C
and $^{10}$B in Ref. \cite{r:NBOFermi}. Here we present the same dependence 
for $^{10}$Li. For the other studied nuclei we choose $\hbar\Omega$ 
in the range 15-17 MeV. We keep the same value of $\hbar\Omega$ 
for each isobaric chain.

Let us note that our calculations
do not violate the separation of the center-of-mass and the internal 
relative motion.
In particular, a variation of the parameter $\beta$ introduced in 
Eq. (\ref{phamomegabeta})
does not change the eigenenergies and other characteristics of the 
physical states.
This is so due to the utilization of a complete $N\hbar\Omega$ 
many-nucleon model space and
the triangular two-nucleon model space for deriving the effective 
interaction as well as due to the procedure used 
to derive the effective interaction.

\subsection{A=7 nuclei}

In Figs. \ref{fig:He7}, \ref{fig:Li7}, \ref{fig:Be7}, and \ref{fig:B7} 
we present the experimental and calculated excitation spectra 
of $^7$He, $^7$Li, $^7$Be, and $^7$B, respectively.  
Their ground-state properties are summarized in the first part of 
Table \ref{tab:tab1}.
The calculations were performed in the $6\hbar\Omega$ model space.
An harmonic-oscillator frequency of $\hbar\Omega=17$ MeV was used.
As discussed earlier in this section, we employed the additional
parameter $k_Q$, introduced in Ref. \cite{r:NB96}, and set its 
value to $k_Q=0.8$.
Note that for $^7$Li and $^7$Be, the dimension in the m-scheme reaches
663,527. It is the largest dimension in the present study.
We should mention that larger matrix dimensions have been used
in other shell-model studies \cite{Poves}, although our calculations 
include more single-particle states. It should still be 
feasible to further extend our no-core calculations to higher
dimensions than presented here.

In general, good agreement with experiment is found for
both the ground-state characteristics as well as the low-lying 
excited states. We observe underbinding for $^7$He and $^7$B.
Note that for an isospin invariant interaction these states would be
degenerate with the $T=3/2$ isopin states of $^7$Li or $^7$Be.
So this underbinding is equivalent to too much spread in the excitation
spectrum. This is a common feature in the no-core shell-model
calculations, which diminishes as the model-space size increases.
Note in Figs. \ref{fig:Li7} and \ref{fig:Be7} the correct 
ordering of excited states in most cases. Also
the magnetic moment of $^7$Li is nicely reproduced. The radii and 
the quadrupole moments are typically smaller in absolute value 
in our calculations. We used bare nucleon charges, so
there is still need for E2 effective charges despite our large
model-space size. By examining the calculated quadrupole matrix
elements, we can deduce that effective charges of 
$e^{\rm p}_{\rm eff}=1.18e$ and $e^{\rm n}_{\rm eff}=0.18e$ are 
needed to obtain the experimental quadrupole moment of $^7$Li.
We note that these effective charges are significantly smaller
than the standard effective charges, 
$e^{\rm p}_{\rm eff}=1.5e$ and  $e^{\rm n}_{\rm eff}=0.5e$,
typically employed in $0\hbar\Omega$ shell-model calculations.

The $^7$Li calculation can be compared to the previously published
$4\hbar\Omega$ no-core calculation of Ref. \cite{ZBVHS}. 
Apart from the larger model-space size in the present calculation,
we also take into account more realistically
the isospin breaking 
and derive the effective interaction in a different way.
The results are not dramatically different. However, the present 
$6\hbar\Omega$
calculation provides a better overall agreement with the experiment.

Our results can also be compared to the recent Green's-function 
Monte Carlo
(GFMC) and Variational Monte Carlo (VMC) calculations \cite{GFMC}. 
It should be noted that the GFMC calculations
are qualitatively different. No effective interaction is used and the 
aim is to obtain the exact many-body solutions.
As the VMC is an upper bound variational calculation, it is
more appropriate to compare the present shell-model calculations
with the GFMC, which includes all the statistically sampled correlations
in the system. Because the GFMC calculations are much 
more involved than VMC,
they have not been carried out for all excited states. 
We should also mention that the GFMC and VMC calculations were done
with the Argonne V18 interaction. 
As both the Argonne V18 interaction and the Reid93 interaction
(that we used) describe 
the NN scattering data more-or-less equally well, 
a different choice of potential should
not cause significant difference in the obtained results.   
In addition, a real three-body
interaction was included in the GFMC and VMC, unlike in our calculations.
Still, it is interesting to note
some similarities between our shell-model calculations
and the VMC and GFMC results for $A=7$. Our $^7$Li calculation shows correct
level ordering for the lowest five states, while the sixth and seventh
states are interchaged in comparison to the experiment. The same feature
is found in the VMC calculation. Also our higher excited states
have energies typically too large, similar to the VMC and GFMC.
On the other hand, the excitation spectrum of $^7$He, obtained in our
calculation, has energies about two times higher than those of the VMC,
although the level ordering is the same.
In addition, in both approaches a larger decrease in the calculated 
binding energies is observed for 
isobars with higher ground-state isospin than is observed experimentally.

\subsection{A=8 nuclei}

In Figs. \ref{fig:He8}, \ref{fig:Li8}, \ref{fig:Be8}, and \ref{fig:B8} 
we present the experimental and calculated excitation spectra of 
$^8$He, $^8$Li, $^8$Be, and $^8$B, respectively. 
Their ground-state properties are summarized in Table \ref{tab:tab1}.
The calculations were performed in a model space of up to $4\hbar\Omega$ 
excitations relative to the unperturbed ground-state configuration. 
An harmonic-oscillator
frequency of $\hbar\Omega=17$ MeV was used. As explained earlier in this
section, the two-body effective interaction was evaluated using 
$N_{\rm max}=8$. We note that the same effective interaction was used
for all the $A=8$ isobars. 

Like in the case of the $A=7$ nuclei, we obtain a good description
of the ground state properties as well as of the low-lying excitation
spectra. In particular, for $^8$Be we have excellent agreement with
the experiment for all positive-parity states below the excitation
energy of 20 MeV. We note that the $T=0, 1$ $J=2^+, 1^+, 3^+$ doublets
show significant isospin mixing compared to other calculated states. 
In our calculations, the lower state always has the $T=1$ component 
dominant. We note that electromagnetic properties of the 16.6 and 16.9 MeV 
$2^+$ doublet in $^8$Be were recently analyzed \cite{Brae95}. The doublet
has almost equal admixtures of $T=0$ and $T=1$ components. 
In Table \ref{tab:tab2}
we compare the experimentally extracted isoscalar and isovector 
electromagnetic transition rates from the doublet with those obtained 
in our shell-model calculation. Our results can also be compared with other 
shell-model calculations as presented in Table III of Ref. \cite{Brae95}.
In those calculations, phenomenological effective interactions of Refs. 
\cite{CK65,K74,HWG88} were employed. 
Our calculation provides excellent agreement with
experiment for the M1 properties and, unlike the other shell-model
calculations used for the analysis, gives the positive sign of the 
isoscalar-isovector matrix element ratio in agreement with experiment.
Also, unlike the other shell-model calculations discussed, we obtained
stronger isoscalar than isovector transition strength for the transition
to the 0+ state.  We note that bare nucleon charges were employed in 
our calculations and that the E2 transitions as well as the quadrupole
moments are typically underestimated.  This indicates a need for
effective E2 charges despite the large model space used. In order
to reproduce the $^8$Li and $^8$B quadrupole moments, we need average
effective
charges of $e^{\rm p}_{\rm eff}=1.3e$ and $e^{\rm n}_{\rm eff}=0.3e$
for this isobaric chain, 
with smaller values for $^8$Li than for $^8$B.

Some excited states of the $A=8$ nuclei have also been calculated using the
VMC approach \cite{GFMC}. Our calculation gives similar results for
the the lowest $2^+ 0$ and $4^+ 0$ states of $^8$Be. On the other hand, 
we get higher excitation energies for the $1^+ 1$ and $3^+ 1$ 
states of $^8$Li, and in particular for the $4^+ 1$ state of $^8$Li 
as well as all the excited states of $^8$He. The largest difference
between our results and the VMC results is in the position of $0^+$
states of $^8$He and $^8$Li. In the VMC calculations their excitation 
energy is significantly lower than that obtained in the shell-model 
calculations. 
As the VMC gives an upper bound, the GFMC result would be lower still.
The $^8$He nucleus is a weakly bound system, where scattering 
to the continuum
will play an important role in the structure of higher-lying states.
Because the harmonic-oscillator basis employed 
in our shell-model calculations
has incorrect asymptotics for the single-particle wave-functions,
we would not expect our calculation to describe well
the higher-lying states in weakly bound systems.

The magnetic moments obtained in our shell-model calculations for 
the $A=7$ and $A=8$ nuclei are in some
cases in a better agreement with the experiment than those obtained
by the VMC. On the other hand, we obtain smaller quadrupole moments
compared to both the experiment and the VMC. Also the rms proton radii
are about 10\% smaller than those obtained in the VMC calculations.

\subsection{A=9 nuclei}

The experimental and calculated negative-parity excitation spectra of  
the $^9$He, $^9$Li, $^9$Be, $^9$B, and $^9$C nuclei are presented 
in Figs. \ref{fig:He9}, \ref{fig:Li9}, \ref{fig:Be9}, \ref{fig:B9},
and \ref{fig:C9}, respectively. 
Their ground state properties are shown in Table \ref{tab:tab1}.
The calculations were performed in the $4\hbar\Omega$ relative to the
unperturbed ground-state configuration model space. We used an
harmonic-oscillator frequency of $\hbar\Omega=16$ MeV. 
As for all the nuclei
with $A\geq 8$, the two-body effective interaction was evaluated using 
$N_{\rm max}=8$ with no additional free parameter. The same effective
interaction was employed for all the $A=9$ nuclei. 

There are few experimental data available, in particular for the
$^9$He, $^9$Li, and $^9$C nuclides. In addition, there are 
no GFMC or VMC calculations, with which we can make a comparison. 
In general, we obtain a correct level ordering
for the lowest states. Also the known magnetic moments are reasonably
reproduced. As in the case of other isobaric chains, the calculated 
absolute values of the quadrupole moments are smaller than the experimental
ones. 
To reproduce the $^9$Li and $^9$Be quadrupole moments, we would need 
average effective charges of $e^{\rm p}_{\rm eff}=1.25e$ 
and $e^{\rm n}_{\rm eff}=0.25e$ for this isobaric chain, 
with smaller values for $^9$Li than for $^9$Be.

Our predictions for the experimentally unknown magnetic moments, 
quadrupole moments
and point-proton rms radii for $^9$He, $^9$B, and $^9$C are given
in Table \ref{tab:tab1}.

\subsection{A=10 nuclei}

We performed extensive calculations for $A=10$ nuclei in order
to evaluate the isospin-mixing correction of the 
$^{10}$C$\rightarrow ^{10}$B Fermi transition. Those calculations
were published in Ref. \cite{r:NBOFermi}. Here we complement the
published results with the calculation for $^{10}$He and $^{10}$Li.
We used the same effective interactions as in Ref. \cite{r:NBOFermi}.
The calculations were performed in the $4\hbar\Omega$ model space.
In Figs. \ref{fig:Li10} we present the dependence of the
$^{10}$Li spectra on the harmonic-oscillator frequency for
$\hbar\Omega=14, 15.5$, and 17 MeV. 
Despite the recent new measurements of $^{10}$Li properties \cite{BZ97},
a controversy on the spin-parity assignment of the ground state
of this nucleus remains. Our calculation prefers $2^+ 2$
as the lowest positive-parity state for all the choices 
of $\hbar\Omega$. 
For $\hbar\Omega=14$ MeV, the $1^+, 2^+$ doublet becomes
almost degenerate, however. We should note that, as discussed 
above for the $^8$He calculation, the shell-model single-particle
wave-functions have incorrect asymptotics. The shell-model
approach is, therefore, not quite suitable for the description
of weakly bound  states or resonances.
A similar dependence on $\hbar\Omega$ for $^{10}$B 
was studied in Ref. \cite{r:NBOFermi}.
There we found a sensitivity of the ground state to $\hbar\Omega$.
Only for $\hbar\Omega=17$ MeV did we obtain the correct ground-state
spin $3^+ 0$, while for $\hbar\Omega=14$ and 15.5 MeV the calculated
ground state was $1^+ 0$. 
In Fig. \ref{fig:Be10} we present the spectrum of $^{10}$Be
obtained in the $4\hbar\Omega$ model space with $\hbar\Omega=15.5$ MeV. 
As discussed in Ref. \cite{r:NBOFermi}, the excited $0^+ 1$ state,
assumed to be dominated by a $2\hbar\Omega$ configuration, is not
obtained below 10 MeV in our calculation. It is likely that our 
$4\hbar\Omega$ model space is not large enough to give 
the right description of such a state.

In Table \ref{tab:tab1}, we present the experimental and calculated 
ground-state properties of the $A=10$ nuclei. Part of the results overlap
with those presented in Table I of Ref. \cite{r:NBOFermi}.
As explained above,
the calculated $^{10}$B $3^+ 0$ for $\hbar\Omega=15.5$ MeV is the 
first-excited state, lying 0.17 MeV above the $1^+ 0$. 
This is the only case
in our study, where the incorrect ground-state spin is obtained. The right
level ordering is recovered for this nucleus, however, 
when $\hbar\Omega$ is increased to, e.g., 17 MeV.
The results presented in Table \ref{tab:tab1} were evaluated
using bare nucleon charges. In order
to reproduce the $^{10}$B experimental quadrupole moment, 
we need effective charges of $e^{\rm p}_{\rm eff}=1.25e$
and $e^{\rm n}_{\rm eff}=0.25e$.

 From the binding energy results given in Table \ref{tab:tab1},
we can deduce the energy splitting between isospin-analog states.
In most cases our calculated splitting is larger than the 
corresponding experimental splitting, though the difference does not 
exceed about 10\%. Mainly the Coulomb energy is responsible for the
isospin-analog-state splitting. As our calculated proton radii are
typically smaller than the experimental ones, we obtain stronger
Coulomb splitting. We discussed this point in Ref. \cite{r:NBOFermi}
and pointed out the importance of the correct value of the proton
radius in order to get the correct isopin-analog-state splitting.

\subsection{A=11 nuclei}

For $A=11$ nuclei we performed calculations only for the $^{11}$Li
and $^{11}$Be. In the last part of Table \ref{tab:tab1} we show
the ground-state properties of $^{11}$Li and the lowest
negative-parity state of $^{11}$Be. In Fig. \ref{fig:Be11} 
we present the experimental and calculated negative-parity spectra 
of $^{11}$Be.
The calculations were performed in the $4\hbar\Omega$ model space
using $\hbar\Omega=15$ MeV.

We note that the ground state of $^{11}$Be is a positive-parity
state $\textstyle{\frac{1}{2}^+}$. We also performed calculations
for the positive parity states in the $3\hbar\Omega$ model space.
We got a correct level ordering for the lowest positive-parity states.
However, those states were shifted with respect to the negative-parity
states by 5.56 MeV, so that the calculated ground state 
has a negative parity.
It should be realized that the relative position of the positive- and
negative-parity states depends on the size of the respective model spaces.
It is quite likely that a $5\hbar\Omega$ model-space calculation would
result in a positive-parity ground state. 

Our $4\hbar\Omega$ model space is insufficient for
reproducing the halo properties of the $^{11}$Li nucleus. This is seen,
in particular,
in a smaller calculated point-proton rms radius as well as a smaller absolute
value of the quadrupole moment compared with the experiment.
In order
to reproduce the $^{11}$Li experimental quadrupole moment, 
we need effective charges of $e^{\rm p}_{\rm eff}=1.27e$
and $e^{\rm n}_{\rm eff}=0.27e$.
On the other hand,
we easily obtain the correct ground-state spin, a reasonable binding energy 
as well as the magnetic moment.

We note that the rms point-proton radii obtained in our calculations
are smaller than the experimental ones with the largest discrepancies
for $^{8,9,11}$Li and  $^{8}$B, e.g., nuclei far from the stability.
One should remark, however, that the experimental extraction of the
rms radii is model dependent.

\section{Conclusions}
\label{sec4}

We have performed large-basis no-core shell-model calculations
for selected $0p$-shell nuclei with $A=7 - 11$. We used two-body
effective interactions derived from the Reid93 nucleon-nucleon
potential with the isospin breaking taken into account. We were
able to reproduce most of the characteristics of the ground states
as well as the correct ordering of the lowest excited states.
As discussed in detail in Section III, our no-core shell-model 
approach has only a very limited number of freely 
adjustable parameters, such as the harmonic-oscillator 
frequency and the size of the model space. 
The calculations were performed
in the $6\hbar\Omega$ and $4\hbar\Omega$ model spaces for the $A=7$ and
$A=8-11$ nuclei, respectively. Our results show, that the 
multi-configuration shell-model approach combined with the use of 
microscopic effective interactions is capable of a good qualitative
and quantitative description of the $0p$-shell nuclei.

It is feasible to extend the present $4\hbar\Omega$ calculations 
to heavier $0p$-shell nuclei as well. For those nuclei slightly
higher m-scheme dimensions than in the present calculations
will have to be dealt with. The present results can be used
for calculating electromagnetic and weak properties and scattering 
characteristics as well as other applications,
like, e.g., $^7$Be$+p \leftrightarrow ^8$B nuclear vertex constant.
With regard to the effective-interaction theory, it is desirable to eliminate
some of the free parameters still present in the calculations
by using three-body effective interactions. 
In particular, we should be able to eliminate the treatment 
of the two-nucleon model-space size
prameter $N_{\rm max}$ as an adjustable parameter, or alternatively 
the use of the parameter $k_Q$ employed in our $6\hbar\Omega$ 
calculations, in this way. Also, the use of the three-body effective
interaction should weaken the dependence on the harmonic-oscillator
frequency. We are presently investigating this aspect in four-nucleon
shell-model calculations. A complete elimination of the dependence
of the results on the harmonic-oscillator frequency will hardly
be achieved, however, in the model-spaces of only a few $\hbar\Omega$
above the unperturbed ground-state configuration,
although it can be greatly weakened. Our investigation of
the three-nucleon shell-model calculations \cite{r:NBthree} supports 
this statement. Most likely,
the three-body effective interaction should not be used directly
as an input into the shell-model calculation, but rather 
it should be utilized for renormalizing
the two-body effective interactions. Work in this direction is under way.

\acknowledgements{
We thank R. Wiringa for useful comments. One of us (BRB) would like to thank
Hans Weidenm\"uller and the Max-Planck Institut f\"ur Kernphysik,
Heidelberg,
and Peter von Brentano and the Institut f\"ur Kernphysik, K\"oln,
Germany, for their hospitality and partial support during the latter stages 
of this work.
This work was supported by the NSF grant No. PHY96-05192.
P.N. also acknowledges partial support from the grant of the 
Grant Agency of the Czech Republic 202/96/1562. 
}

\begin{figure}
\caption{The experimental and calculated excitation spectra of $^{7}$He.
The results corresponding to the model-space size
of $6\hbar\Omega$ relative to the unperturbed 
ground-state configuration are presented.
An harmonic-oscillator energy of $\hbar\Omega=17$ MeV was used.
}
\label{fig:He7}
\end{figure}

\begin{figure}
\caption{The experimental and calculated excitation spectra of $^{7}$Li.
The results corresponding to the model-space size
of $6\hbar\Omega$ relative to the unperturbed 
ground-state configuration are presented.
An harmonic-oscillator energy of $\hbar\Omega=17$ MeV was used.
}
\label{fig:Li7}
\end{figure}

\begin{figure}
\caption{The experimental and calculated excitation spectra of $^{7}$Be.
The results corresponding to the model-space size
of $6\hbar\Omega$ relative to the unperturbed 
ground-state configuration are presented.
An harmonic-oscillator energy of $\hbar\Omega=17$ MeV was used.
}
\label{fig:Be7}
\end{figure}

\begin{figure}
\caption{The experimental and calculated excitation spectra of $^{7}$B.
The results corresponding to the model-space size
of $6\hbar\Omega$ relative to the unperturbed 
ground-state configuration are presented.
An harmonic-oscillator energy of $\hbar\Omega=17$ MeV was used.
}
\label{fig:B7}
\end{figure}

\begin{figure}
\caption{The experimental and calculated excitation spectra of $^{8}$He.
The results corresponding to the model-space size
of $4\hbar\Omega$ relative to the unperturbed 
ground-state configuration are presented.
An harmonic-oscillator energy of $\hbar\Omega=17$ MeV was used.
}
\label{fig:He8}
\end{figure}

\begin{figure}
\caption{The experimental and calculated excitation spectra of $^{8}$Li.
The results corresponding to the model-space size
of $4\hbar\Omega$ relative to the unperturbed 
ground-state configuration are presented.
An harmonic-oscillator energy of $\hbar\Omega=17$ MeV was used.
}
\label{fig:Li8}
\end{figure}

\begin{figure}
\caption{The experimental and calculated excitation spectra of $^{8}$Be.
The results corresponding to the model-space size
of $4\hbar\Omega$ relative to the unperturbed 
ground-state configuration are presented.
An harmonic-oscillator energy of $\hbar\Omega=17$ MeV was used.
}
\label{fig:Be8}
\end{figure}

\begin{figure}
\caption{The experimental and calculated excitation spectra of $^{8}$B.
The results corresponding to the model-space size
of $4\hbar\Omega$ relative to the unperturbed 
ground-state configuration are presented.
An harmonic-oscillator energy of $\hbar\Omega=17$ MeV was used.
}
\label{fig:B8}
\end{figure}

\begin{figure}
\caption{The experimental and calculated excitation spectra of $^{9}$He.
The results corresponding to the model-space size
of $4\hbar\Omega$ relative to the unperturbed 
ground-state configuration are presented.
An harmonic-oscillator energy of $\hbar\Omega=16$ MeV was used.
}
\label{fig:He9}
\end{figure}

\begin{figure}
\caption{The experimental and calculated excitation spectra of $^{9}$Li.
The results corresponding to the model-space size
of $4\hbar\Omega$ relative to the unperturbed 
ground-state configuration are presented.
An harmonic-oscillator energy of $\hbar\Omega=16$ MeV was used.
}
\label{fig:Li9}
\end{figure}

\begin{figure}
\caption{The experimental and calculated excitation spectra of $^{9}$Be.
The results corresponding to the model-space size
of $4\hbar\Omega$ relative to the unperturbed 
ground-state configuration are presented.
An harmonic-oscillator energy of $\hbar\Omega=16$ MeV was used.
}
\label{fig:Be9}
\end{figure}

\begin{figure}
\caption{The experimental and calculated excitation spectra of $^{9}$B.
The results corresponding to the model-space size
of $4\hbar\Omega$ relative to the unperturbed 
ground-state configuration are presented.
An harmonic-oscillator energy of $\hbar\Omega=16$ MeV was used.
}
\label{fig:B9}
\end{figure}

\begin{figure}
\caption{The experimental and calculated excitation spectra of $^{9}$C.
The results corresponding to the model-space size
of $4\hbar\Omega$ relative to the unperturbed 
ground-state configuration are presented.
An harmonic-oscillator energy of $\hbar\Omega=16$ MeV was used.
}
\label{fig:C9}
\end{figure}

\begin{figure}
\caption{The dependence of the spectra of $^{10}$Li on
the harmonic-oscillator energy for $\hbar\Omega=14$ MeV,
15.5 MeV, and 17 MeV, respectively.
The results corresponding to the model-space size
of $4\hbar\Omega$ relative to the unperturbed 
ground-state configuration are presented.
}
\label{fig:Li10}
\end{figure}

\begin{figure}
\caption{The experimental and calculated excitation spectra of $^{10}$Be.
The results corresponding to the model-space size
of $4\hbar\Omega$ relative to the unperturbed 
ground-state configuration are presented.
An harmonic-oscillator energy of $\hbar\Omega=15.5$ MeV was used.
}
\label{fig:Be10}
\end{figure}

\begin{figure}
\caption{The experimental and calculated excitation spectra of $^{11}$Be.
The results corresponding to the model-space size
of $4\hbar\Omega$ relative to the unperturbed 
negative-parity ground-state configuration are presented.
An harmonic-oscillator energy of $\hbar\Omega=15$ MeV was used.
Note that the ground state of this nucleus has positive parity.
The energy of the calculated $\textstyle{\frac{1}{2}^-}$ 
state is set equal to the lowest
experimental negative-parity state.}
\label{fig:Be11}
\end{figure}

\begin{table}
\begin{tabular}{c|cc|cc|cc|cc|cc}
Isotope &  $^{7}$He & &$^{7}$Li& & $^{7}$Be & & $^{7}$B & & & \\
A=7 & Calc. & Exp. & Calc. & Exp. & Calc. & Exp. & Calc. & Exp. & &\\
\hline
$J^{\pi} T$ & $\textstyle{\frac{3}{2}^- \frac{3}{2}}$ & 
 $\textstyle{(\frac{3}{2})^- \frac{3}{2}}$ &
 $\textstyle{\frac{3}{2}^- \frac{1}{2}}$ &
 $\textstyle{\frac{3}{2}^- \frac{1}{2}}$ &   
 $\textstyle{\frac{3}{2}^- \frac{1}{2}}$ &
 $\textstyle{\frac{3}{2}^- \frac{1}{2}}$ &
$\textstyle{\frac{3}{2}^- \frac{3}{2}}$ & 
 $\textstyle{(\frac{3}{2})^- \frac{3}{2}}$ & & \\   
$E_B$ & 26.926 & 28.82(3) & 39.270 & 39.245 & 37.632 
& 37.600 &22.466 & 24.72&&\\
$\mu$ & -1.166 &       & +2.994 & +3.256 & -1.132 &      & +2.921&      &&\\
$Q$   & +0.471 &       & -2.710 & -4.00(6)  & -4.631 &      & +4.338&    &&\\
$\sqrt{\langle r_p^2\rangle}$& 1.692&&2.045 &2.27(2)&2.216&2.36(2)
&2.472& &&\\
\hline
Isotope &  $^{8}$He & &$^{8}$Li& & $^{8}$Be & & $^{8}$B & & & \\
A=8 & Calc. & Exp. & Calc. & Exp. & Calc. & Exp. & Calc. & Exp. & &\\
\hline
$J^{\pi} T$ &$0^+2$&$0^+2$&$2^+1$&$2^+1$&$0^+0$&$0^+0$&$2^+1$&$2^+1$&&\\  
$E_B$ & 25.426 & 31.408 & 36.859 & 41.277 & 52.486 & 56.500 
&33.033 & 37.738&&\\
$\mu$ & -      &  -    & +1.419 & +1.654 & -      & -    & +1.240&  1.036&&\\
$Q$   & -      &  -    & +2.208 & +3.11(5)& -     & -    & +4.000
&(+)6.83(21)&&\\
$\sqrt{\langle r_p^2\rangle}$& 1.684&1.76(3)&1.941&2.26(2)&2.071&
&2.188&2.45(5) &&\\
\hline
Isotope &  $^{9}$He & &$^{9}$Li& & $^{9}$Be & & $^{9}$B & &$^{9}$C &  \\
A=9 & Calc. & Exp. & Calc. & Exp. & Calc. & Exp. & Calc. & Exp. 
& Calc.& Exp.\\
\hline
$J^{\pi} T$ & $\textstyle{\frac{1}{2}^- \frac{5}{2}}$ & 
 ? &
 $\textstyle{\frac{3}{2}^- \frac{3}{2}}$ &
 $\textstyle{\frac{3}{2}^- \frac{3}{2}}$ &   
 $\textstyle{\frac{3}{2}^- \frac{1}{2}}$ &
 $\textstyle{\frac{3}{2}^- \frac{1}{2}}$ &
 $\textstyle{\frac{3}{2}^- \frac{1}{2}}$ &
 $\textstyle{\frac{3}{2}^- \frac{1}{2}}$ &
$\textstyle{\frac{3}{2}^- \frac{3}{2}}$ & 
 $\textstyle{(\frac{3}{2}^-) \frac{3}{2}}$  \\   
$E_B$ & 23.048 & 30.26(6)& 40.827&45.341 &55.194 &58.165 &53.082 & 56.314&
34.343& 39.034 \\
$\mu$ & +0.656 &       & +2.940 & 3.439 & -1.066 &-1.178& +2.865&      &
-0.981&1.391\\
$Q$   & - & - & -2.085 & -2.74(10)  & +3.245 & +5.29(4)& +2.582&      &
-2.591&\\
$\sqrt{\langle r_p^2\rangle}$& 1.740&&1.946 &2.18(2)&2.063&2.34(1)&2.169& &
2.259& 2.48(3)\\
\hline
Isotope &  $^{10}$He & &$^{10}$Li& & $^{10}$Be & & $^{10}$B & &$^{10}$C &  \\
A=10 & Calc. & Exp. & Calc. & Exp. & Calc. & Exp. & Calc. & Exp. 
& Calc.& Exp.\\
\hline
$J^{\pi} T$ &$0^+2$& ? &$2^+2$&$?\; 2$&$0^+1$&$0^+1$&$3^+0$&$3^+0$&
$0^+1$&$0^+1$\\  
$E_B$ & 22.653 &30.34(7)&40.923 & 45.316 & 63.024 & 64.977&62.607 & 64.751&
58.194&60.321\\
$\mu$ & -      &  -    & +3.105 &   & -      & -    & +1.850& +1.801&
- & - \\
$Q$   & -      &  -    & -2.046 &   & -      & -    & +5.643 & +8.472&
- & - \\
$\sqrt{\langle r_p^2\rangle}$& 1.786&&1.958 &&2.051&2.24(8)&2.127&2.30(12)&
2.214& 2.31(3)\\
\hline
Isotope &&&  $^{11}$Li & &$^{11}$Be& & & & &   \\
A=11 &&& Calc. & Exp. & Calc. & Exp. & & & &  \\
\hline
$J^{\pi} T$ & & &$\textstyle{\frac{3}{2}^- \frac{5}{2}}$ & 
 $\textstyle{\frac{3}{2}^- \frac{5}{2}}$ & 
 $\textstyle{\frac{1}{2}^- \frac{3}{2}}$ &
 $\textstyle{\frac{1}{2}^- \frac{3}{2}}$ &   
 & & &   \\   
$E_B$ &&& 43.686 & 45.64(3)& 65.124&65.481 &&&&  \\
$\mu$ &&& +3.601& 3.668& +0.802 &  & & & &  \\
$Q$   &&& -2.301 &-3.12(45)& - & - &  & & &  \\
$\sqrt{\langle r_p^2\rangle}$&&& 1.986&2.88(11)&2.061 &&&&& 
\end{tabular}
\caption{Experimental and calculated ground-state spins and parities;
binding energies, in MeV; magnetic moments, in $\mu_{\rm N}$;
quadrupole moments, in $e$ fm$^2$; and the point proton rms radii, 
in fm, of the nuclei studied. 
The results correspond to the 6$\hbar\Omega$ for $A=7$ and
4$\hbar\Omega$ for $A\geq 8$ calculations, respectively.
The harmonic-oscillator parameter was taken to be $\hbar\Omega=17$ MeV
for $A=7,8$, 16 MeV for $A=9$, 15.5 MeV for $A=10$, and 15 MeV for $A=11$,
respectively. 
The effective interaction used was derived from the Reid 93 
nucleon-nucleon potential. The Coulomb interaction and isospin breaking
in $T=1$ partial waves was taken into account. 
The same effective interaction was used
for all nuclei of a given isobaric chain characterized by $A$.
Bare nucleon charges were used in the calculations.   
The experimental values are taken from Refs. 
\protect\cite{r:AS88,VDV87,r:T88,T93,r:OT96}.}
\label{tab:tab1}
\end{table}

\begin{table}
\begin{tabular}{cccc}
Final state & Observable & Calc. & Expt.  \\
\hline
$2^+$(3.0 MeV) & B(M1, $IV$) & 0.086 & $0.091\pm 0.006$  \\
$2^+$(3.0 MeV) & B(M1, $IS$) & $3\times 10^{-4}$ 
& $(2\pm 2)\times 10^{-4}$  \\
$2^+$(3.0 MeV) & $\epsilon$  & +0.057 & $+0.06\pm 0.02$  \\
$2^+$(3.0 MeV) & B(GT)       &  0.018 & 0.031  \\
$2^+$(3.0 MeV) & B(E2, $IV$) & $2\times 10^{-4}$ 
& $(1\pm 6)\times 10^{-3}$  \\ 
$2^+$(3.0 MeV) & B(E2, $IS$) & 0.063 & $0.30\pm 0.12$  \\
$0^+$(0.0 MeV) & B(E2, $IV$) & 0.024 & $0.00\pm 0.03$ 
or $0.14\pm 0.03$ \\
$0^+$(0.0 MeV) & B(E2, $IS$) & 0.028 & $0.14\pm 0.03$ 
or $0.00\pm 0.03$ \\
$2^+$(3.0 MeV) & $\delta_1$  & +0.005 & $+0.01\pm 0.03$  \\
$2^+$(3.0 MeV) & $\delta_0$  & +0.099 & $+0.22\pm 0.04$  \\
\end{tabular}
\caption{Experimental and calculated properties
of transitions from the 16 MeV $2^+$ doublet
in $^8$Be. Reduced transition strengths are given in $e^2$fm$^4$
for E2 and in $\mu_{\rm N}^2$ for M1. The observable $\epsilon$
is defined as a ratio of the isoscalar and the isovector M1 matrix 
elements. The mixing ratios $\delta_0$ ($\delta_1$) are computed
using the E2 isoscalar (isovector) matrix element and the
isovector M1 matrix element. 
Bare nucleon charges were used in the calculations.  
The experimental values are taken from Ref. 
\protect\cite{Brae95}.}
\label{tab:tab2}
\end{table}

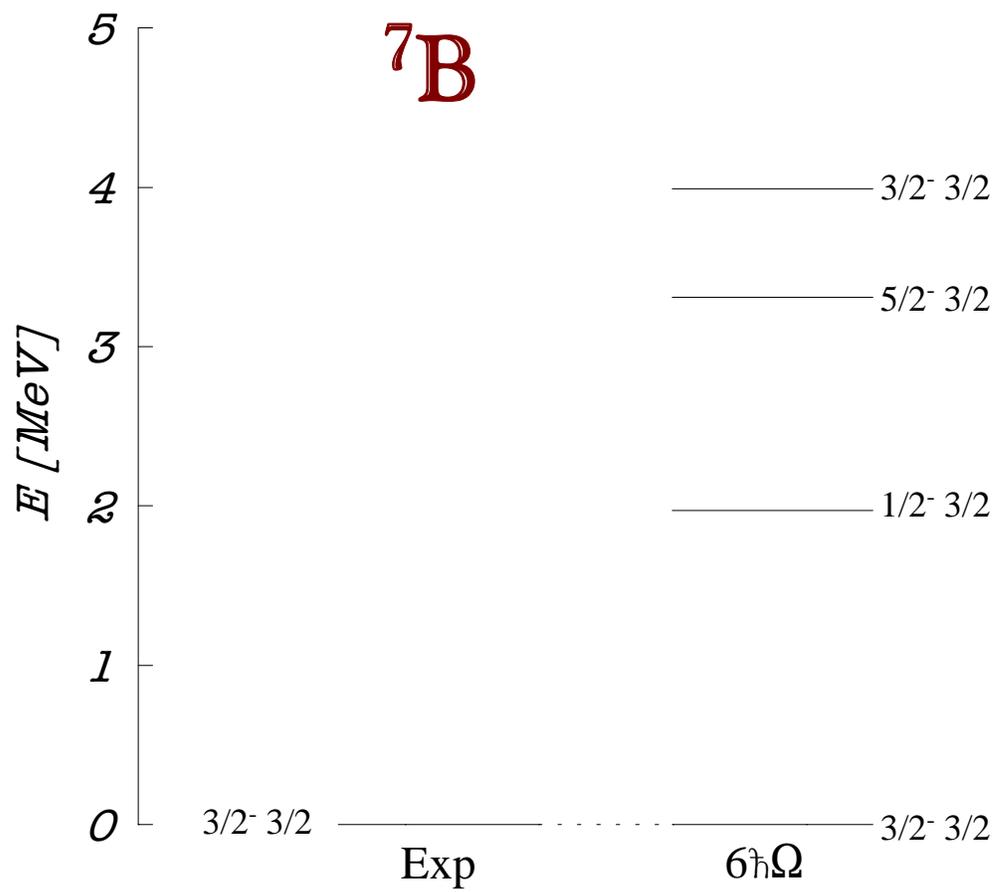
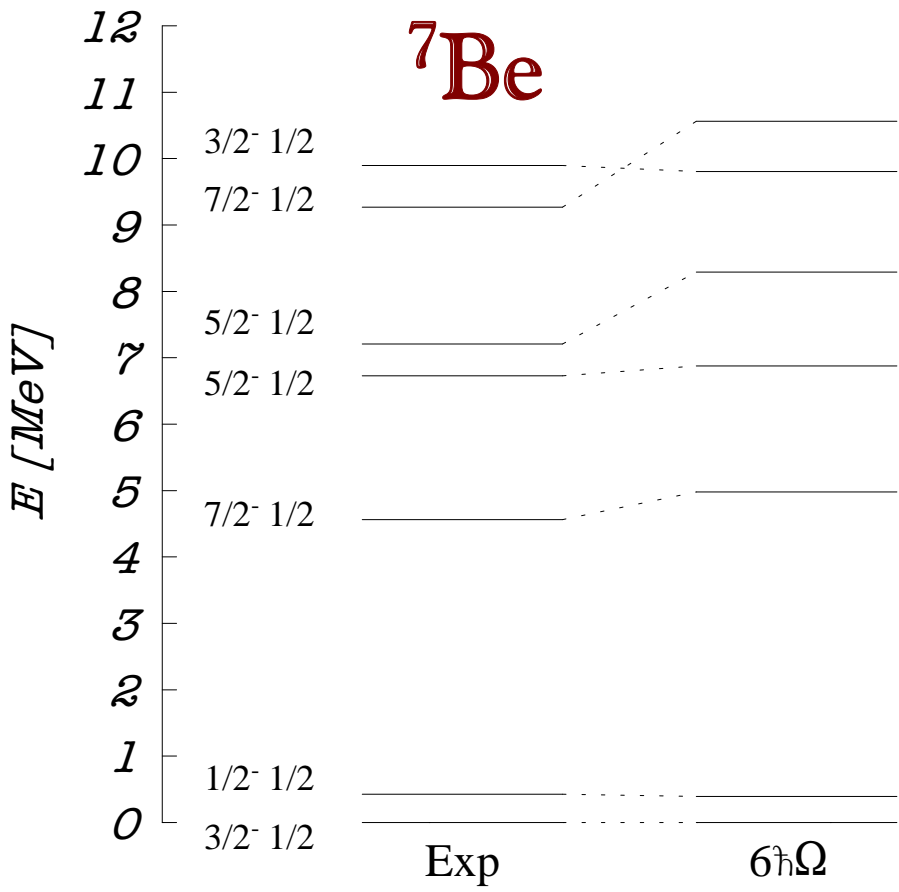
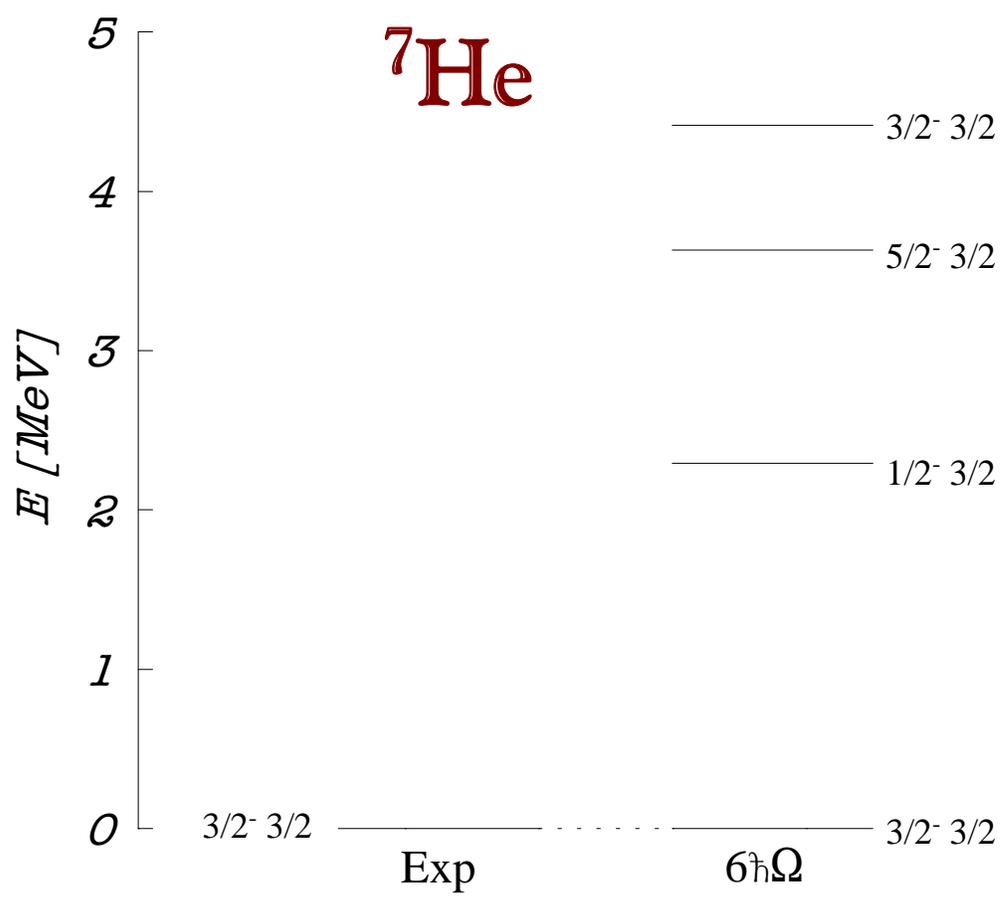
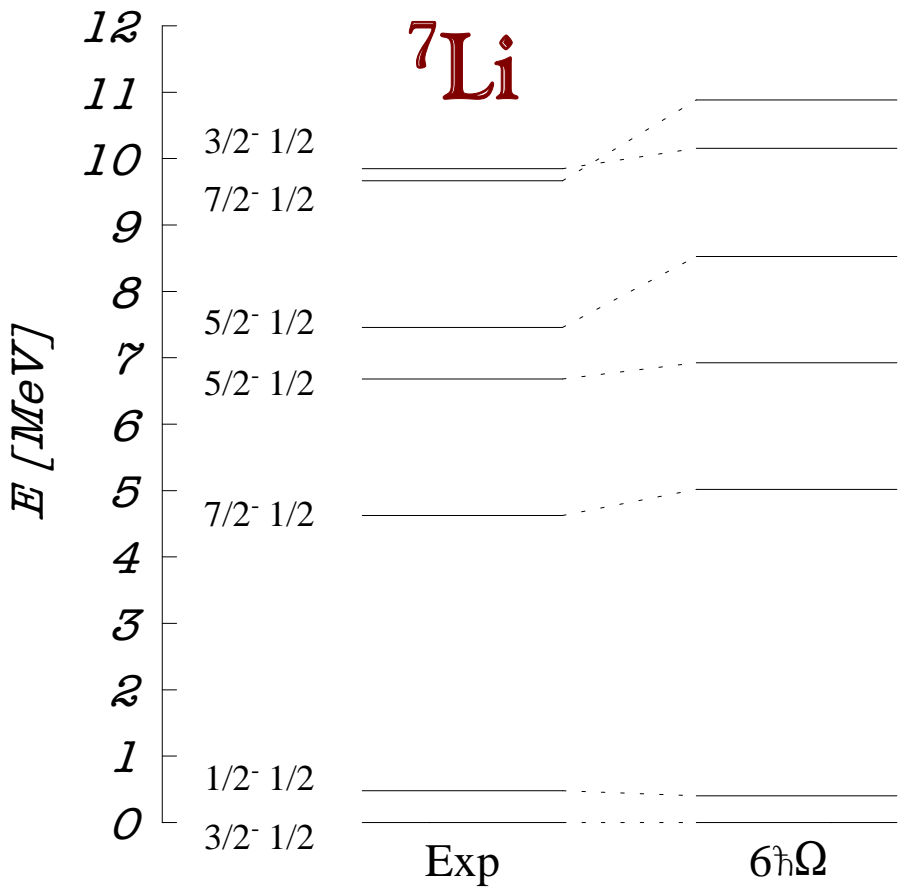
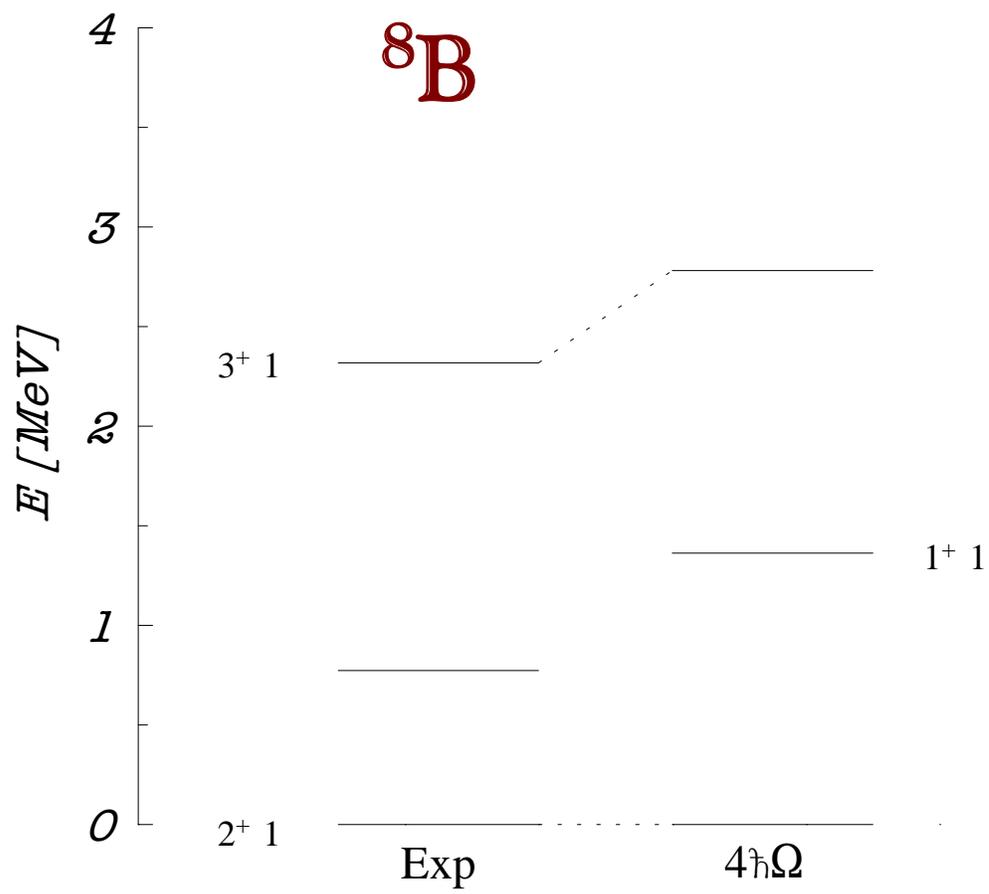
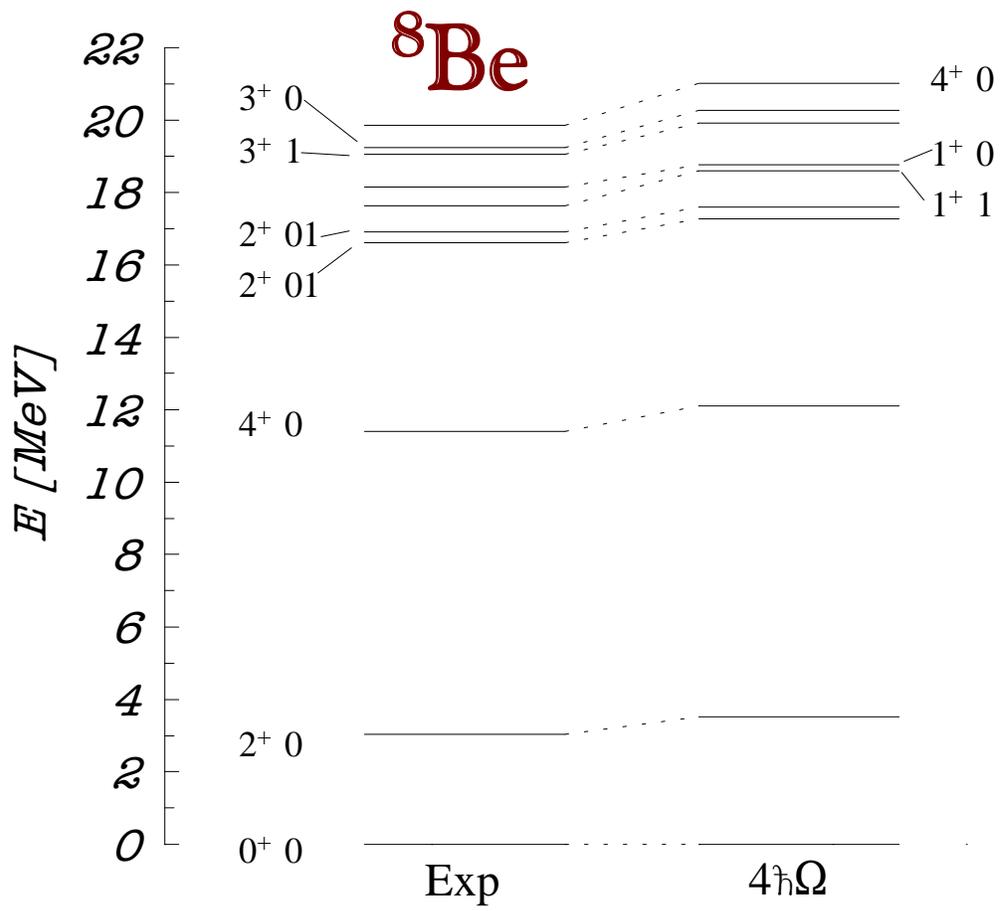
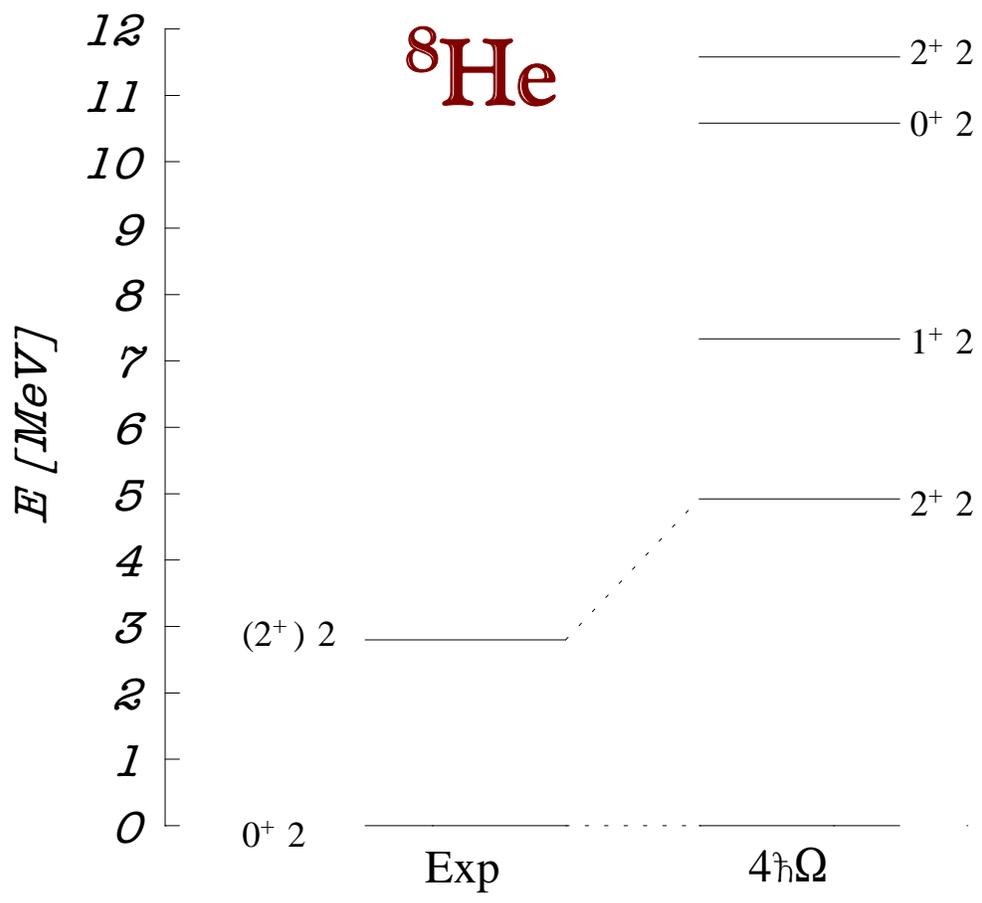
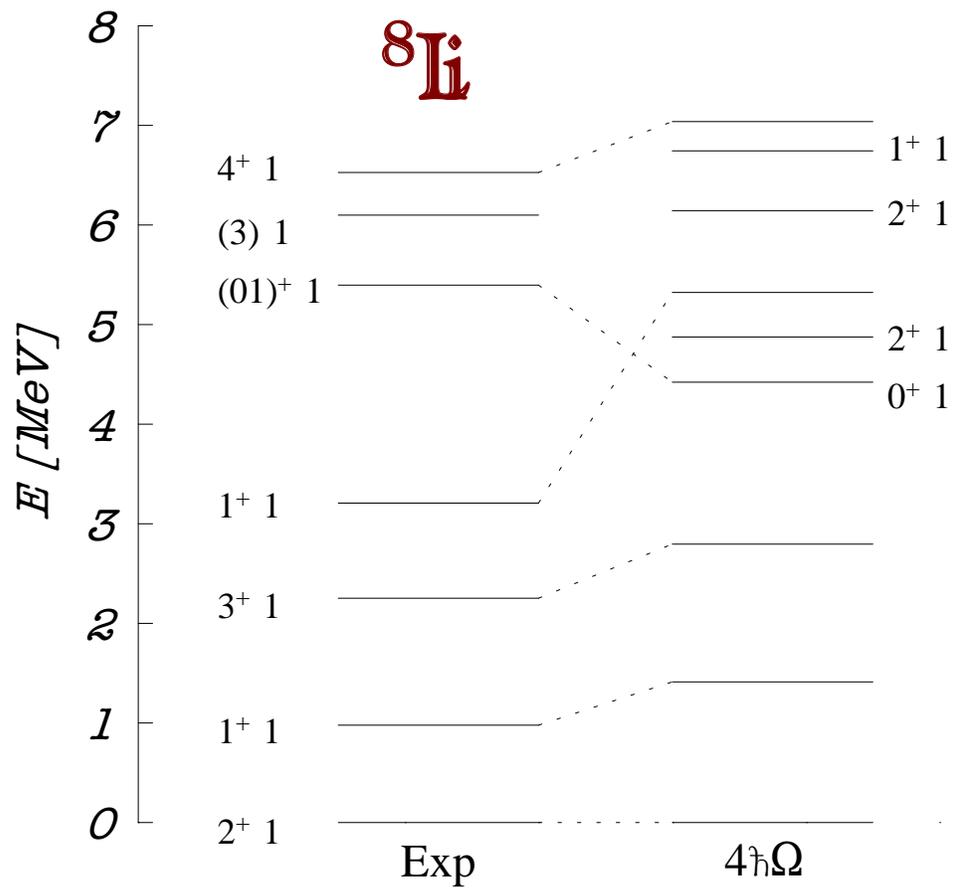
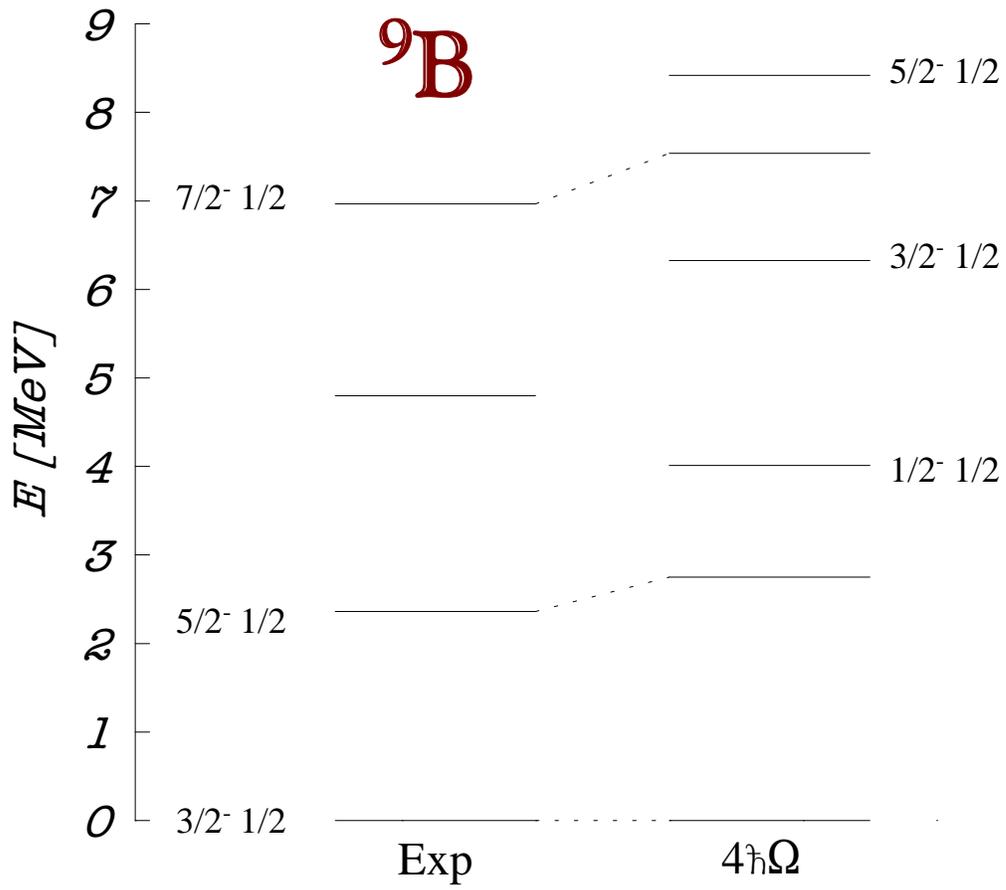
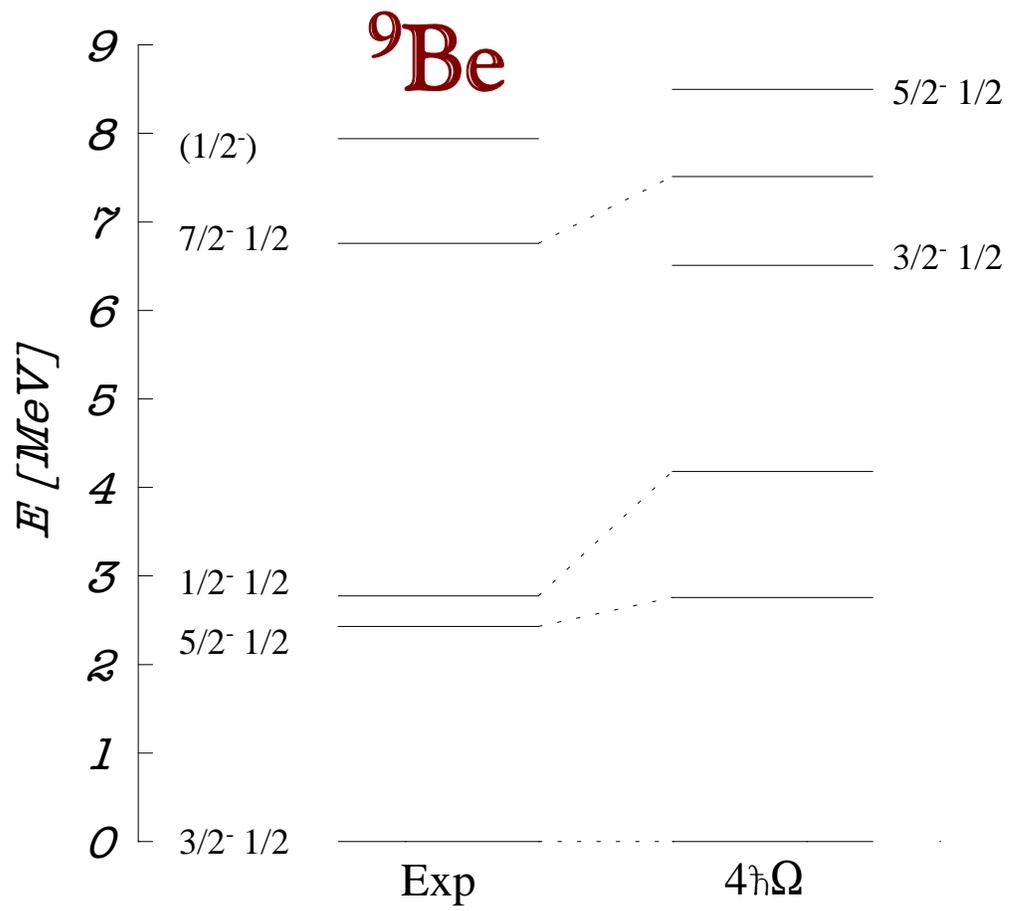
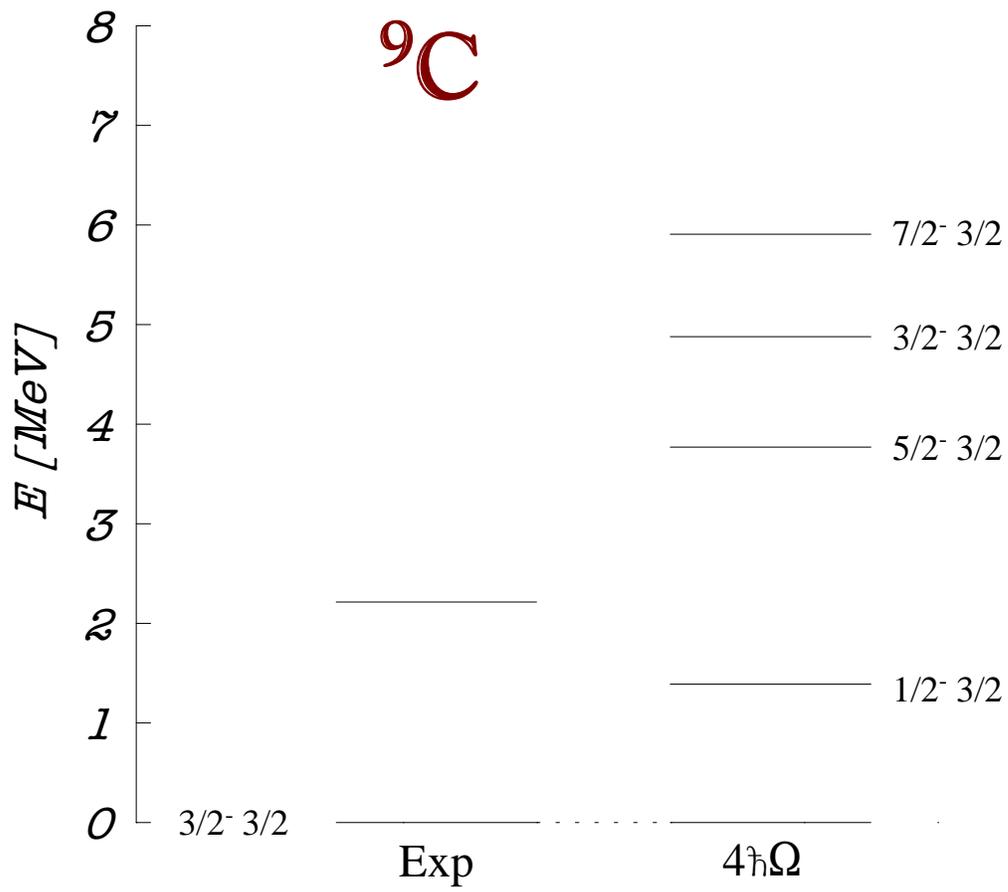
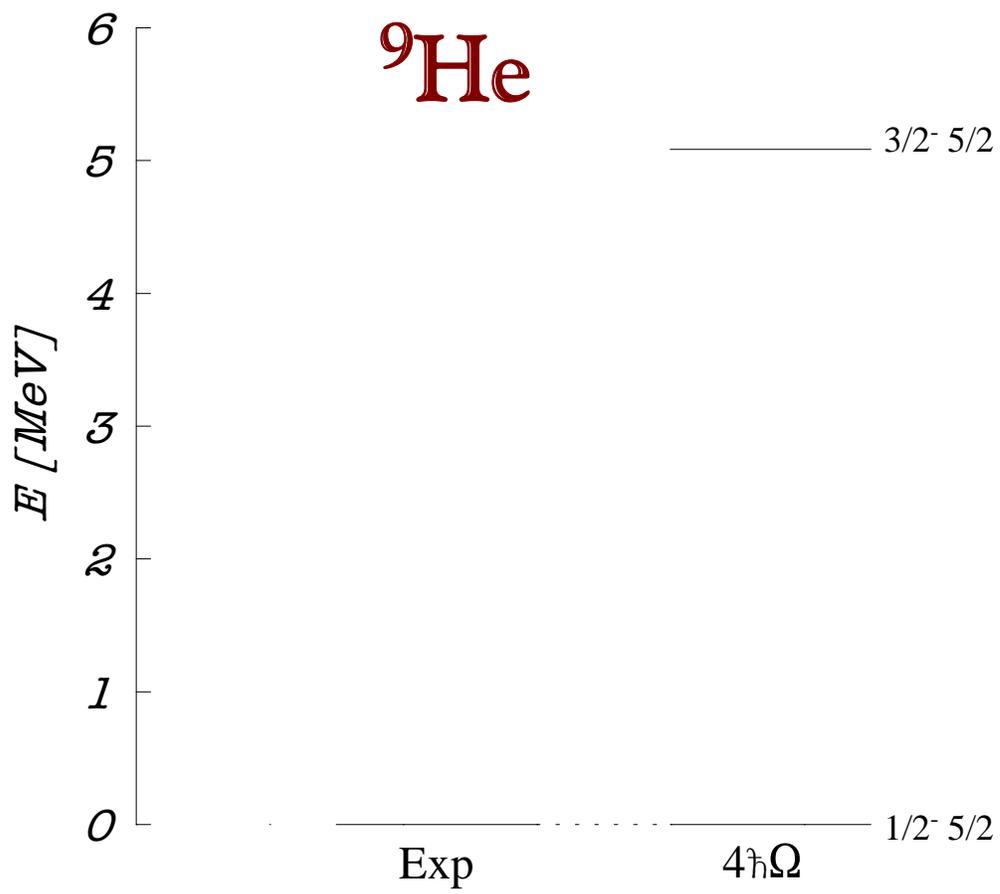
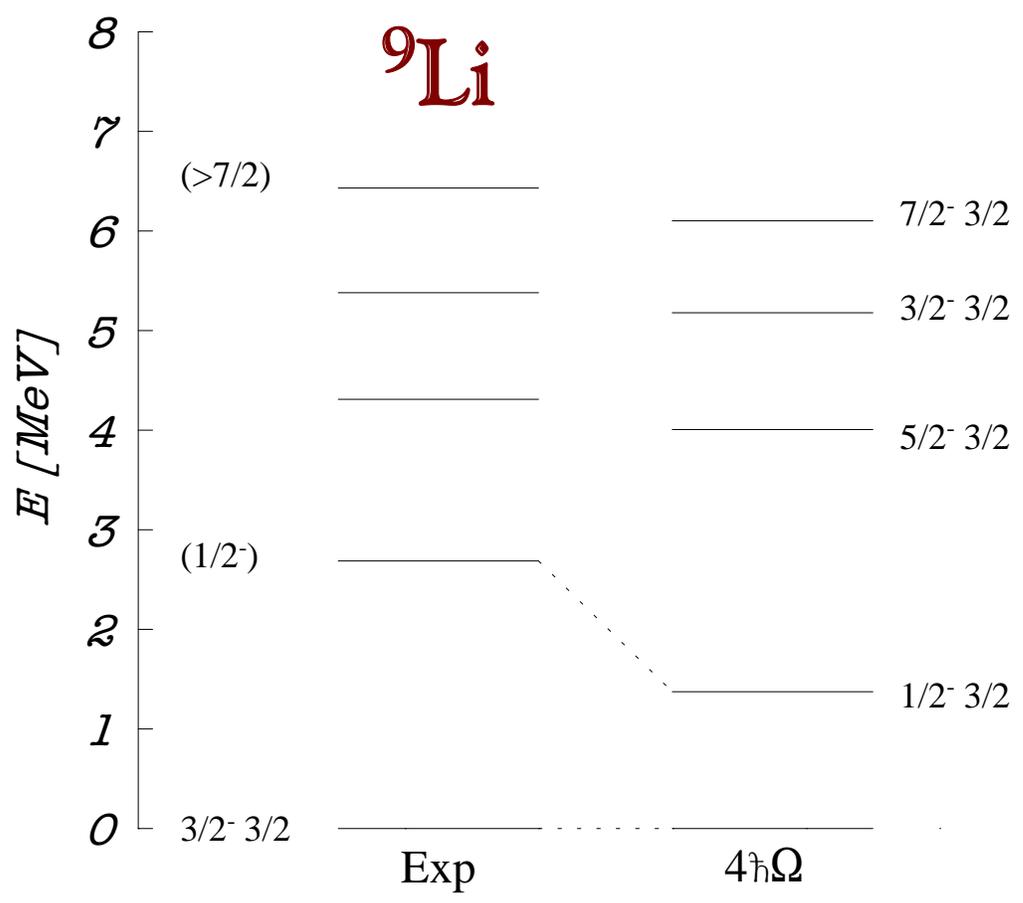
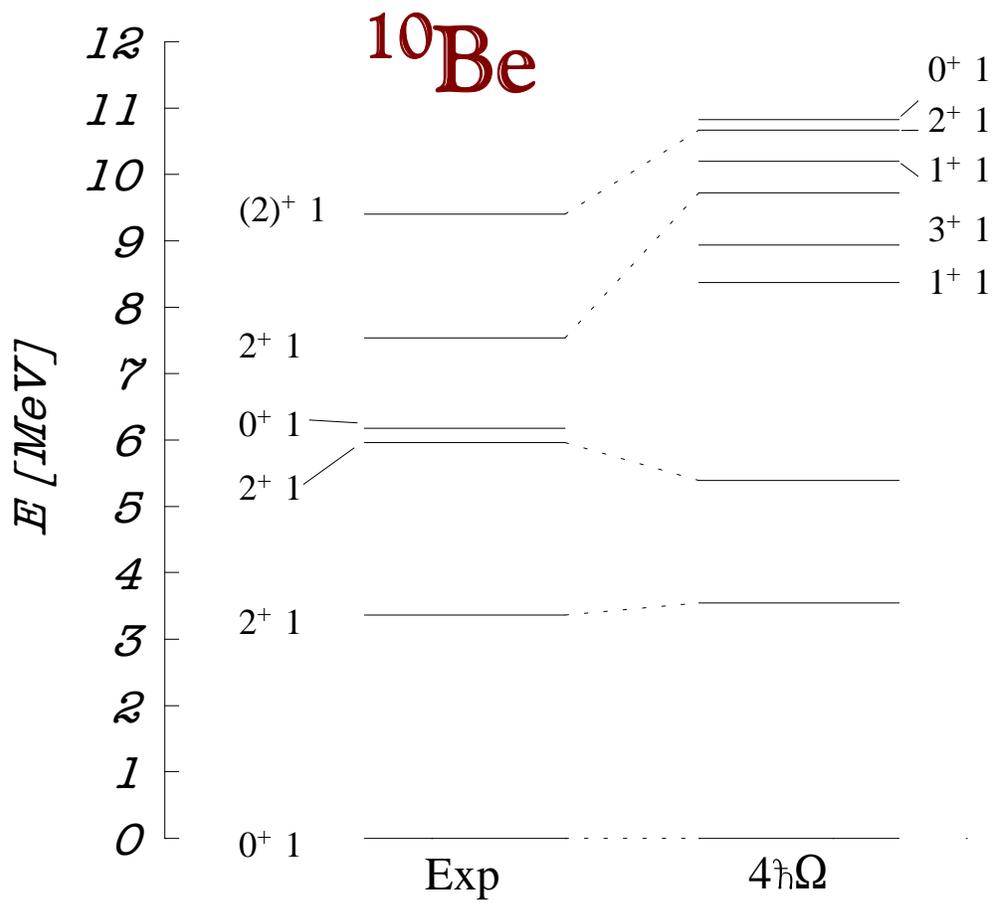
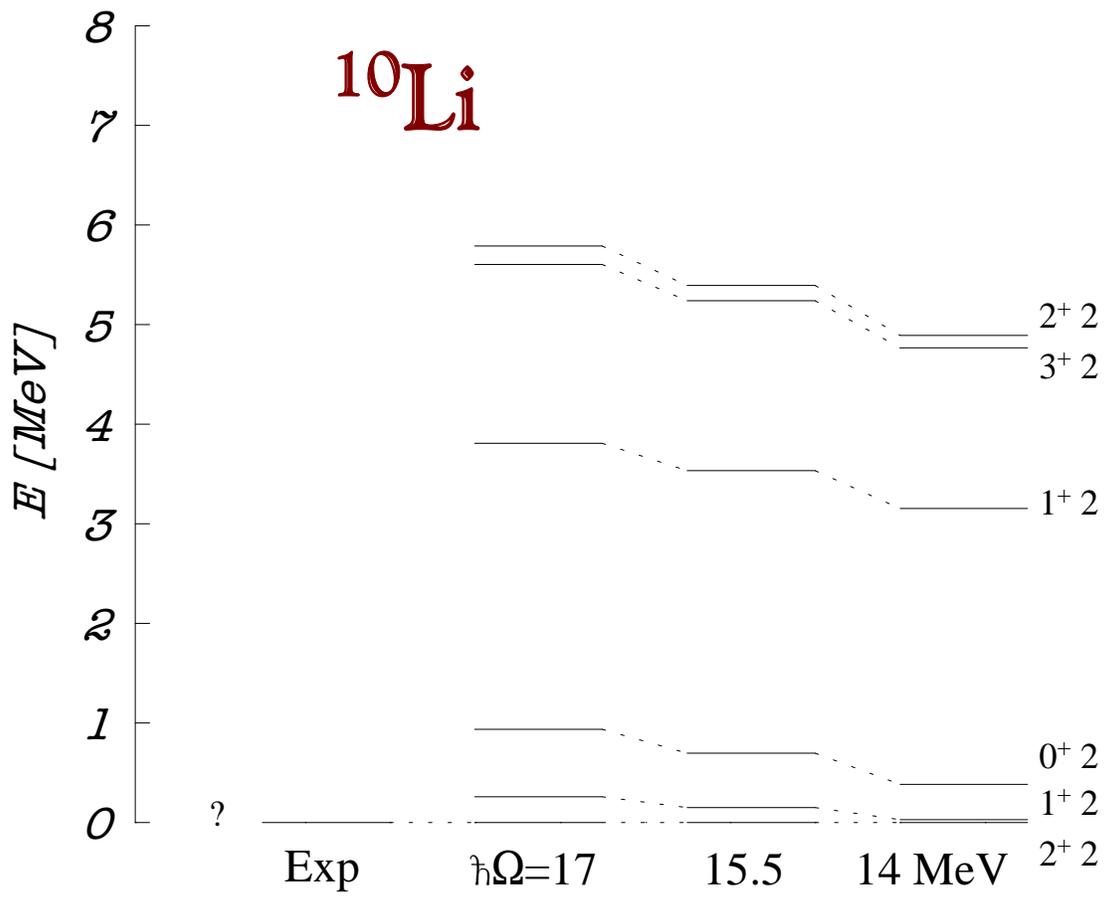
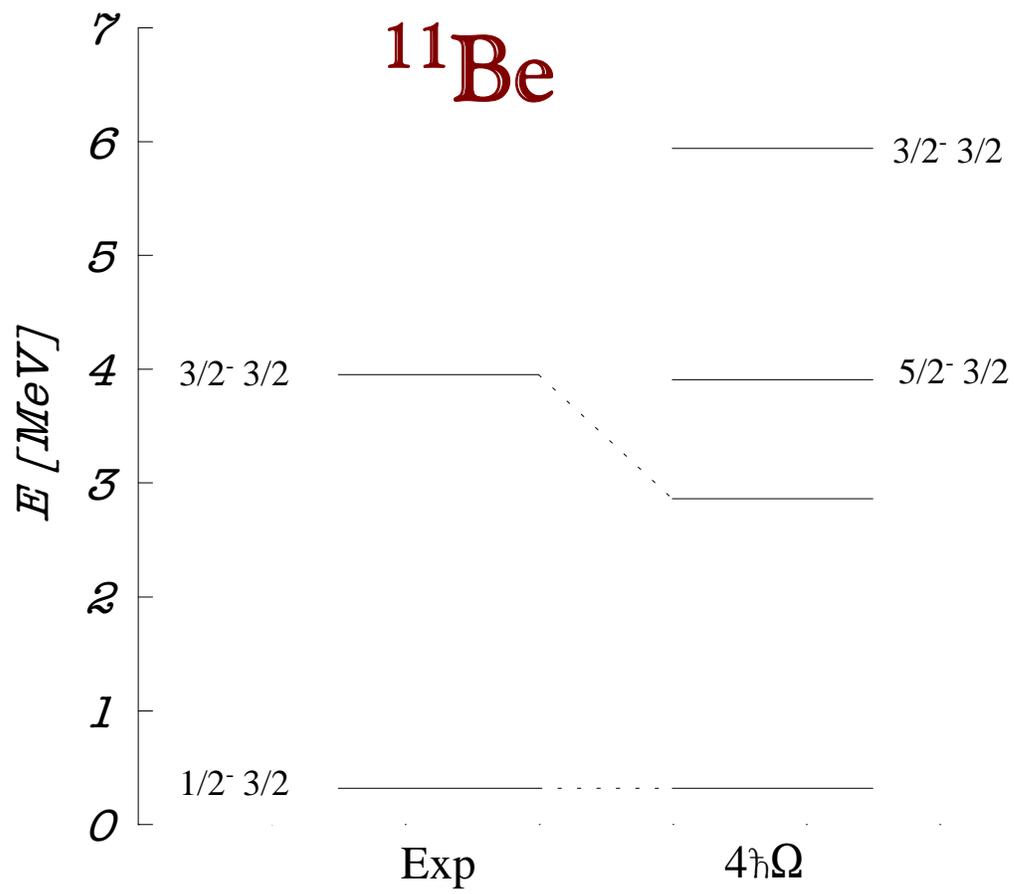
\end{document}